# Multimode image memory based on a cold atomic ensemble

Dong-Sheng Ding, Jing-Hui Wu, Zhi-Yuan Zhou, Yang Liu, Bao-Sen Shi,[*] Xu-Bo Zou,[1] and Guang-Can Guo

*Key Laboratory of Quantum Information, University of Science and Technology of China, Hefei 230026, China*

Corresponding author: [*]drshi@ustc.edu.cn

[1]xbz@ustc.edu.cn

**Abstract**

Quantum memory is one of the key components constituting a quantum network. An important step towards the successful development of such a network is the storage of single photons. Encoding photons in high-dimensional photonic states can significantly increase the network's information-carrying capacity. Furthermore, quantum memories that are able to store multiple optical modes offer advantages over single-mode memories, both in terms of speed and capacity. However, a practical implementation of such a scheme for storing and retrieving multiple images at the single-photon level has not yet been achieved. Here, we provide the first experimental evidence that two spatial probe fields imprinted with a real image each can be stored and retrieved with good visibility and similarity at the single-photon level in single $^{85}$Rb cold atomic ensemble, making use of electromagnetically induced transparency. Our results are very promising towards the realization of a high-dimensional quantum network in the future.

## I. INTRODUCTION

A significant goal in the field of quantum communication is the development of a quantum network through which users can exchange quantum information at will. Such a network would consist of spatially separated devices used to store and manipulate quantum information, and quantum channels through which different devices would be connected. Photons, which interact weakly with their environment, are robust and efficient carriers of quantum information. Atoms are well suited for precise quantum-state manipulation and long-term storage of quantum information in meta-stable states [1, 2].

Images and other spatial structures have long been a communication medium of choice because of the large amount of information these could carry. Quantum information can also be encoded in the spatial degrees of freedom of a photon, for example, orbital angular momentum [3–5], or transverse momentum and position [6]. In comparison with two-dimensional states, high-dimensional states show many interesting properties. For example, higher-dimensional states enable more efficient quantum-information processing, and large-alphabet quantum key distribution affords an increased flux of information [7]. Besides, a quantum memory capable of controlling the storage and release of a photon encoded in a high-dimensional state is a key component in realizing long-distance high-dimensional quantum communication.

Some progress has been made in storing images in memories, such as those based on a hot atomic ensemble [8, 9] and a doped solid system [10]. In Ref. 8, a transverse image of a 5-bar test pattern was stored and retrieved using a combination of electromagnetically induced transparency (EIT) and four-wave mixing techniques. In Ref. 9, an image of a digit was stored using the EIT technique. In Ref. 10, a digit image was stored using EIT in a cryogenically cooled doped crystal. However, all experiments mentioned before were performed with high intensity light; to date, there have been no reports on the storage and retrieval of an image imprinted on light at the single-photon level. Moreover, quantum memories that are able to store multiple optical images offer advantages in terms of speed and robustness over single-image memories—advantages that in turn can lead to higher efficiency in quantum communication and computation experiments [11–14]. There have been some related experimental progress in this direction, such as storing light with multimode structure in the time domain using the gradient echo memory scheme in a hot atomic ensemble [15] or in the frequency domain using the atomic frequency comb technique in solids doped with rare-earth-metal ions [16], but there have been no reports on the storage of multiple optical images in any domain in single memory. There have been also no reports on the storage of multiple optical images at the single-photon level in any memory system.

In this work, we report on the first experimental demonstration of the simultaneous storage and retrieval of two images imprinted on two different light pulses at the single-photon level through EIT in a cold $^{85}$Rb atomic cloud, where the spatial information is stored in atomic collective spin excitation state and retrieved later. Our work takes an important step towards realizing long-distance high-dimensional quantum communication using atom-based quantum memories. Along with the recent important progress made in the areas of infrared-to-visible wavelength conversion [17] and long-distance fiber transmission of a photon encoded in a high-dimensional state [18], our results could lay the basis for establishing a high-dimensional quantum network in the future.

## II. EXPERIMENTAL RESULTS

EIT could be used to make an opaque medium transparent by means of quantum interference. Meanwhile, optical properties of the medium would be greatly modified. It would significantly enhance the nonlinear susceptibility in the spectral region of induced transparency and also induce steep dispersion. Steep dispersion is necessary in realizing light-slowing and -storage. In a typical storage experiment, the information contained in one degree (frequency, amplitude, phase, and spatial mode) of light is mapped into long-lived atomic coherence by switching off adiabatically the control light. A fraction of the signal pulse remains in the cell before the control field is turned off, which results in an observed signal-leakage that is not affected by the storage operation. Sometime later, the coherence could be read out into an electromagnetic field by turning the control light back on. In our experiments, we used a Λ-type EIT configuration (see Fig. 1), consisting of two ground states, |1> and |2>, and one excited state |3>. The ground states are given by the Zeeman-degenerate levels of the $^{85}$Rb atom ($5S_{1/2}$, F=3, $m_F$=-1 and -3), the excited state corresponds to the level $5P_{1/2}$, F=2 $m_F$=-2. A cigar-shaped atomic cloud of $^{85}$Rb atoms was obtained in a two-dimensional magneto-optical trap (MOT) and served as the memory element in our experiment [19]. The size of the cloud was about $30 \times 2 \times 2$ mm$^3$. The total number of atoms was $9.1 \times 10^8$. The probe and coupling fields from an external-cavity diode laser (DL100, Toptica) had the same wavelength of 795 nm. The coupling field was used to address the transition between states |1> and |2>. The probe field was divided into two probes (probe 1 and probe 2) by a polarization beam splitter, and propagated through the atomic cloud along different directions. The angle between the coupling and probe 1 was 3.3 °, and the angle between probe 1 and probe 2 was 0.45 °. The non-collinear configuration used in the experiment significantly reduces the noise from the coupling scattering. Both probe fields coupled states |2> and |3>, and each was imprinted with a real digital image through a mask of a standard resolution chart (USAF target). The continuous-wave (CW) probe fields were modulated by an acoustic-optical modulator (AOM) to form a sequence of 500-ns long light pulses. These pulses were focused onto the atomic cloud using a lens with 300-mm focal length. The coupling beam was 3 mm in diameter and covered the probe beams completely. By adjusting a quarter-wave plate before the MOT, the coupling and probe fields were assigned opposite circular polarizations. Using another quarter-wave plate after the MOT, the fields were reversed to have orthogonal linear polarizations. The power of the coupling field was 50 μW. The probe powers could be adjusted via an attenuator. The retrieved signals were directed into a photomultiplier tube (Hamamatsu, H10721) for temporal intensity measurement, or a time-resolved camera (CCD, 1024×10, iStar 334T series, Andor) triggered by a synchronization signal generated by a signal generator (AFG 3252) for spatial structure measurement. The total size of the camera sensor was $13.3 \times 13.3 mm^2$. The CCD had a quantum efficiency of approximately 25% at 795-nm wavelength. The mask plane, the center plane of the atomic cloud and the imaging plane of CCD consisted of a 4-f imaging system using two lenses. (See bottom inset of Fig. 1)

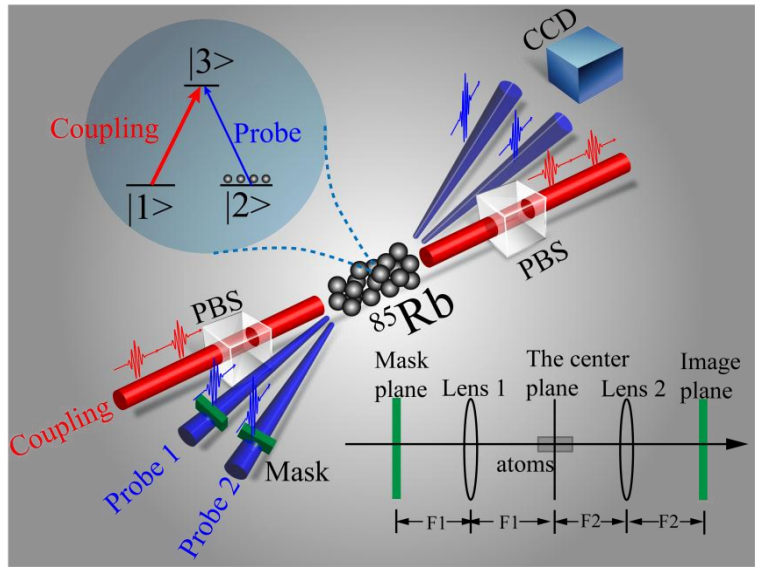

**Fig. 1 (Color online) Experimental setup.** Top inset is the simplified energy diagram and bottom inset is the 4-f system used. The focus lengths of lens 1 and lens 2 are $F_1$=300 mm and $F_2$=500 mm respectively.

We carried out the experiments at a repetition frequency of 1 kHz. The time sequence for the measurement is given in Fig. 2(a). Each iteration consisted of an atomic loading period of 800 μs and an experimental window of 200 μs that accommodated 50 probe and coupling pulses. Therefore, the total number of probe pulses per second was 50,000. The periods of the probe and coupling pulses were 3.536 μs, and the widths of each probe and coupling pulse were 500 ns and 1.61 μs, respectively. To begin, we used two photomultiplier tubes to record the temporal intensities of both probes. Figure 2(b) shows the experimental results. In our system, the back-edge of each coupling pulse wrote the probe into the atomic collective spin excitation; the front-edge of the following pulse was used to read out the stored probe pulse from its spin excitation. We found that such a design was able to significantly improve the strength of the retrieved signals. The first half of each probe pulse transmitted directly (we call this part the 'leakage pulse'), and only the remaining part was stored in the experiment. We calculated the storage efficiencies by comparing the intensities of the retrieved signal and the leakage pulse, which were 0.35 for probe 1 and 0.23 for probe 2 respectively.

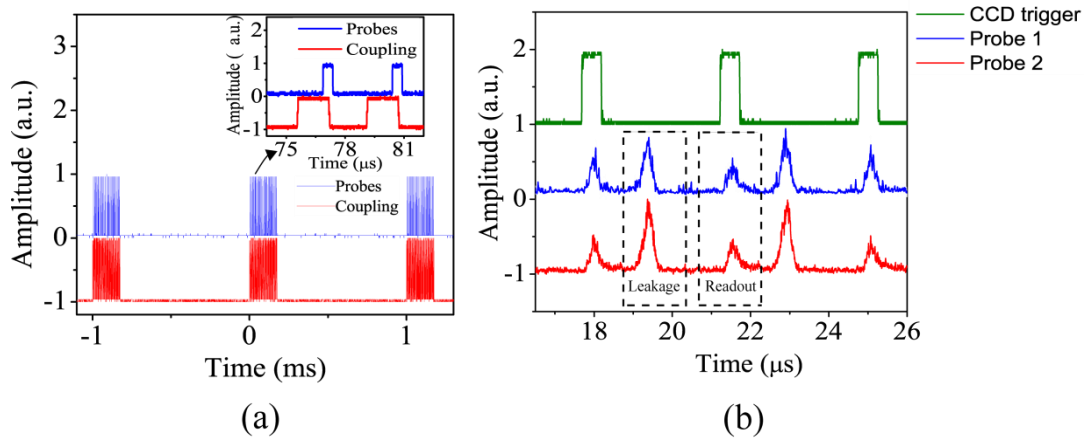

**Fig. 2 (Color online) Timing sequence for the measurement. (a)** Pulses sequence of probe and coupling. Blue (upper): probe fields; red (down): coupling fields. The inset shows two pulses expanded along the x-axis. **(b)** leakage and retrieved pulses of the probe. The blue (middle) and red (down) solid lines show the measured leakage and retrieved signals of probe. The green (upper) line shows the synchronization. signal for triggering the camera.

Next, we used the CCD camera for spatial mode analysis. In this case, the rising edge of a synchronized trigger signal with a 500-ns width was used to trigger the shutter of the high-time-resolution camera. After triggering, the shutter remained open for about 500 ns before closing. To record the retrieved images, the shutter of the camera should keep open while the retrieved signal arrived at CCD camera. By properly adjusting the delay of the synchronization signal (the trigger delay can be adjusted by the camera itself or by using a signal generator), we could make the shutter window and the retrieved signal overlap in time. (The time when the coupling pulse is applied determined when the stored signal was retrieved. The shutter window could be adjusted. Here we set it at 500 ns because the width of probe pulse was 500 ns.) In this way, the retrieved images were recorded by the CCD camera (see Figure 3(b)). If the shutter window did not overlap with the retrieved signal, then no image could be retrieved. Similarly, we could also record the leakage signals (Figure 3(a)). In both Figures 3(a) and (b), the left column shows the image of probe 1 and the right column the image of probe 2. The storage time we set — that is, the time between the leakage pulse and the read-out pulse — was approximately 1.8 μs. Qualitatively, the leakage images and the retrieved images are very similar.

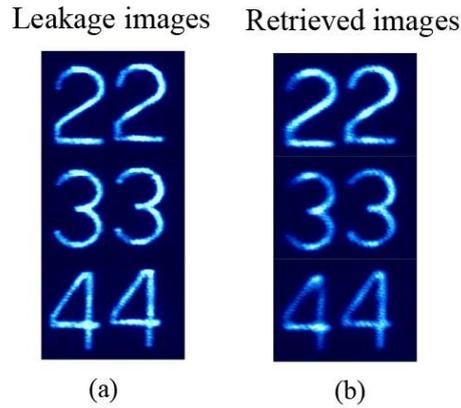

**Fig. 3 (Color online) Multiply images storage (a)** Leakage images imprinted on probes. **(b)** The retrieved images. The imprinted probes with different digits were stored for 1.826 μs, and the CCD camera was exposed to each image for 0.3 s. There were approximately 1000 photons in each probe pulse.

We also investigated cross-talk between the stored images. In this experiment, two different images were stored. We separately considered the storage of probe 1, probe 2, and probe 1+2. The experimental results are displayed in Fig. 4, where, (a),(c),(e) were the leaked images of probe 1, probe 2 and probe 1+2, and (b), (c), (d) were their retrievals. The results clearly show no cross-talk between the two stored images.

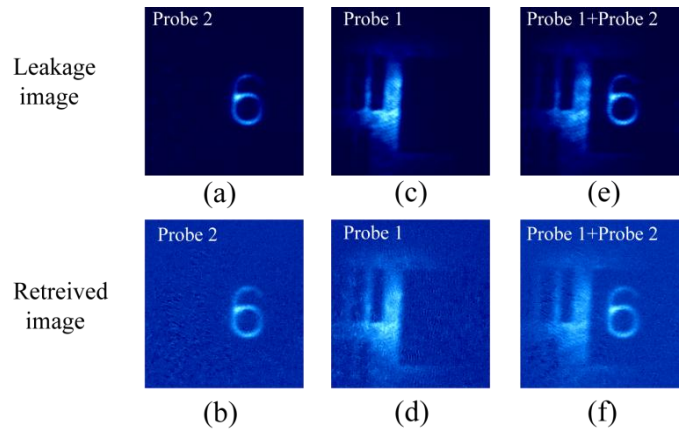

**Fig. 4 (Color online) Storage of two probe fields imprinted with different images.** Pairing (a) and (b), (c) and (d), (e) and (f) are leaked and retrieved images of probe 2, probe 1, and probe 1+2 respectively.

Next, we gradually reduced the intensity of the probe fields with attenuators. In the following experiments, an image of digit "2" was taken as an example. Figure 5(b–g) correspond to the storage of a pulse with the average number of photons of 305, 162, 80, 40, 22, 10, and 5.3 respectively. (Also in these experiments, we only stored half of each probe pulse. The photon number of each probe was estimated as follows: we used a single-photon detector (PerkinElmer SPCM-AQR-15-FC) to count the photon number of each probe per second with no atomic cloud. After taking into account transmission losses to the detector and the efficiency of the detector, we estimated the averaged photon number in each probe pulse. We want to mention that the method used here to determine photon number per pulse is not accurate when the photon number per pulse is large. The SPCM photon counting module used has a dead time of about 50 ns; therefore, the photon number estimated according to measurement should be less than the actual photon number contained in a 500-ns probe pulse. When the photon number per pulse is very low, with the decrease of the photon number in a pulse, the estimated value becomes closer to the actual value. For example, if we assume that there are two photons in a pulse, and these two photons are distributed equally during the square pulse (here, we assume that the probe pulse has a square profile for simplicity), the probability of the second photon arriving at the detector during the detector's dead time is about 0.1 after the first photon has been detected. If we assume that the response time of the detector is smaller than the time interval between these two photons, the estimated photon number in the pulse is about 1.9, close to the actual value. Therefore, the estimation of the photon number per pulse is more accurate when the pulse is very weak, especially for ten or less photons per pulse. ) Even when storing the probe pulses with the average of 1.2 photons, we could still clearly observe the retrieved images (see Figure 5(h)). By plotting the intensity profiles in the vertical direction through the center of the digital '2', we calculated the visibilities of the retrieved images using formula $V=\frac{I_{max}-I_{min}}{I_{max}+I_{min}}$, where $I_{max}$ and $I_{min}$ are the maximal and minimal photon counts along the vertical direction. Figure 6(a) shows the visibilities of the retrieved images with different photon numbers in each probe. The visibility clearly increased with the number of photons in each probe. We also calculated the similarity R of the retrieved images compared with the leakage images using the formula $R=\frac{\sum_m \sum_n A_{mn} B_{mn}}{\sqrt{\sum_m \sum_n A_{mn}^2 \sum_m \sum_n B_{mn}^2}}$, where $A_{mn}$ and $B_{mn}$ are the gray-scale intensities recorded for pixel m, n of the two images to be compared. Figure 6(b) shows the similarities of the retrieved images against the different photon numbers in each probe. For 1.2 photons on average per pulse, the visibilities for the retrieved images for probe 1 and probe 2 were still above 50% and 40%, respectively, and the similarities of the retrieved images were more than 70%, respectively.

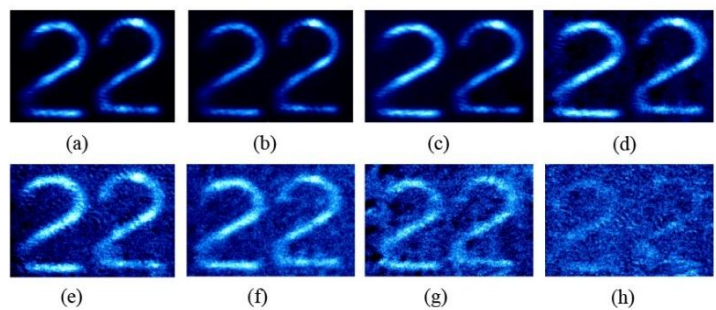

**Fig. 5 (Color online) Retrieved images at low photon number per pulse.** The storage time for all images was 1.826 μs. In each panel, the left image was obtained from probe 1 and the right image from probe 2. In panels (**a**)−(**c**), each image represented the sum of 50 retrieved images; for (**a**) 305, (**b**) 162, and (**c**) 80 photons on average per probe pulse. In panels (**d**)

and (**e**), the average photons were 40 and 22 per probe pulse, respectively, and each image was the sum of 200 retrieved images. In panel (**f**), the average photons were 10 per pulse, and the image was the sum of 500 retrieved images; in panels (**g**) and (**h**), the average photons were 5.3 and 1.2 per probe pulse, respectively, and each image was the sum of 1000 retrieved images. The images shown here were obtained by subtracting the background. (The background was measured by switching off both probes and only keeping the coupling light on.) In each panel, the exposure time of the CCD camera for one retrieved image was 1.0 s.

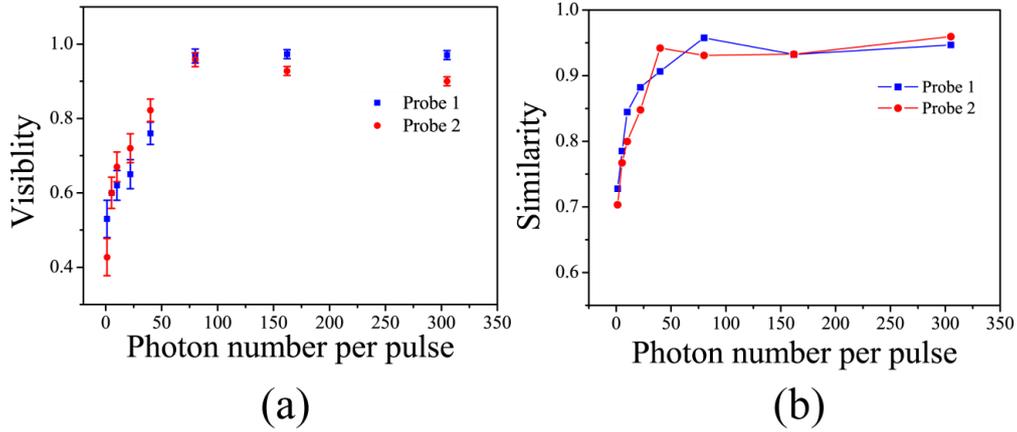

**Fig. 6 (Color online) Visibilities and similarities of the retrieved images.** (**a**) The visibility of the signals (Probe 1: blue square; Probe 2: red circle) versus the photon number per pulse. Error bars represent $\pm$ one standard deviations, and were obtained based on the statistics of the CCD counting events. (**b**) The similarity of the signals versus the photon number per pulse. The blue (square) line corresponded to the images from probe 1; the red (circle) line to images from probe 2. The solid line is given as a visual guide.

## III. DISCUSSIONS AND ANALYSIS

The figures presented in Figure 5 clearly provide the experimental evidence that multiple image memories at the single-photon level can be realized using a cold atomic ensemble; the main features of the image had been preserved during the process. Figure 6 shows that the visibility and the similarity decreased with the reduction of the photon number in a probe pulse. When the photon number of a probe pulse was reduced close to the noise level, the photon counts contributed from the noise approached the counts caused by the probe pulse, so that the signal-to-noise ratio (SNR) was greatly reduced and the image became unclear. Therefore, the noise should be decreased further for a more clear retrieved image. We believe one source of noise is the scattering of the coupling pulse when we switched it on to retrieve the stored images. The ground states used in this experiment were degenerate, therefore, it was quite difficult to reduce the noise caused by coupling as the coupling and the probe pulses had the same wavelength. Fortunately, such noise could be reduced significantly by using a non-collinear EIT configuration, as done in our present experiments. We found that with the increase in angle between the probe and coupling pulses, the noise reduced gradually. In principle, we might further increase the angle between coupling and probe to reduce the influence of this noise source further, but further increasing the angle would reduce the EIT dip and result in a decrease in memory efficiency. Therefore, we had to reach a compromise between angle and achievable efficiency. In our experiment, we had taken a suitable angle by observing the EIT; increasing the angle as far as possible under the EIT condition have no obvious change compared with that at zero angle. Choosing two non-degenerate ground states instead might be a better strategy for suppressing the scattering coupling, because that can further reduce the noise from the coupling pulse using for example a spectrum filter. The second noise source was from the dephasing between two ground states induced by the earth magnetic field and the atomic motion. It not only reduced the SNR, but also shortened the storage time. Another noise source was from our laboratory environment. The very weak lights from the liquid crystal indicators or indicator lamps from the various kinds of equipment would also contribute to the dark counts of the CCD camera. Moreover, the large sensor area of the CCD camera ($13.3 \times 13.3 mm^2$) also recorded more noise photons. From Fig. 7, we see that the edge of the image somehow softened, which was mainly caused by atomic diffusion. In our experiment, the cold atomic ensemble was used, the speed of atomic motion was very small compared with the atom in a hot vapor ensemble, so the diffusion was reduced significantly. Furthermore, a 4-f imaging system was used in our experiment, the Fourier transform of the image, instead of the image itself, was stored in the atomic ensemble. By this configuration, diffusion can be further reduced [8]. Therefore, the clear edge could still be kept for a long time. We emphasize here that the storage time of the probe at the single-photon level could be increased, for example, to over dozens of microseconds, (Fig. 7 and Ref. 20), but with the increase in the storage time, the probability of retrieving a photon became small, hence the noise contributed more counts to the CCD detector, which greatly reduced the SNR and blurred the retrieved image. In the experiments reported here, we kept storage time relatively short to curtail the overall duration of the experiments, so that a relatively high SNR and a clear image could be achieved. In Fig. 5, it seems that probe 2 suffered a loss of sharpness a little more than probe 1; this could be due to slightly different power and beam waists between these two probes. In Fig. 6(a), the visibilities for probe 2 decreased marginally when the photon number per pulse went beyond 80; this could be due to relatively small accumulations of the retrieved images.

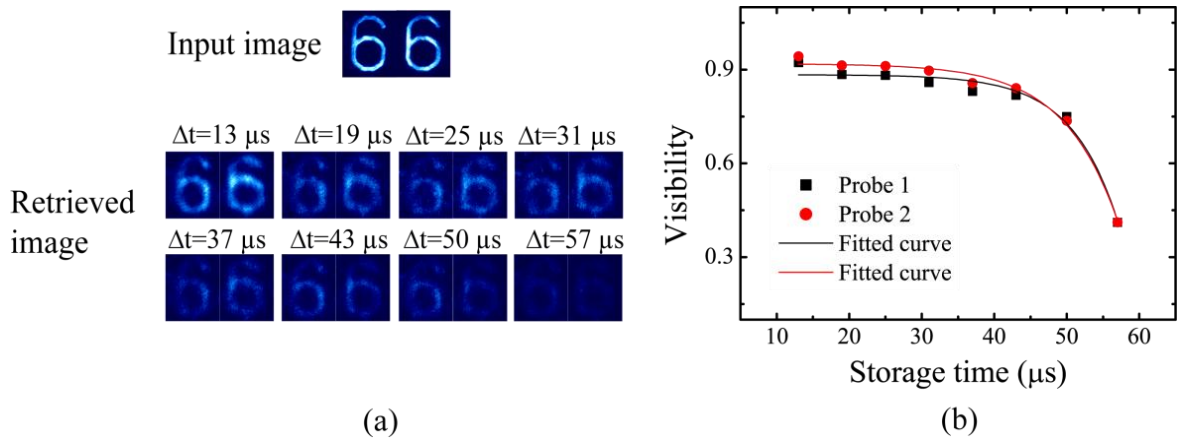

**Fig. 7 (Color online)The retrieved images against the time of the storage.** The CCD camera exposure was 0.3 s for each image. There were approximately 1000 photons in each probe pulse. (b) The intensity signal vs. storage time. The visibility was calculated using $I_{max}$ and $I_{min}$ along the horizontal direction. Experimental data were fitted with exponential functions $y = y_0 + A * E^{(t-t_0)/\tau}$ (where $\tau$ is the decay time, $y_0$, $A$, $t_0$ are the fitted parameters).

## IV. Conclusion

In summary, we demonstrated experimentally that multiply probes with an optical image each at the single-photon level could be stored in and retrieved from an atomic ensemble using EIT. All images, including those obtained with light pulses containing only 1.2 photons per pulse, were characterized by high similarities and good visibilities. Our work takes an important step toward realizing long-distance high-dimensional quantum communication with atomic-based quantum memories.

*During the submission of this manuscript, we noted one report on spatial mode storage in a gradient echo memory [21], in which one spatial optical mode is considered and the memory does not work at the single-photon level, and another report on storing two different images [22] using the same technique as in Ref. 21.*

**Acknowledgements**

This work was supported by the National Natural Science Foundation of China (Grant Nos. 11174271, 61275115, 10874171), the National Fundamental Research Program of China (Grant No. 2011CB00200), and the Innovation Fund from CAS, Program for NCET.